\documentclass[usenatbib,useAMS]{mn2e}
\usepackage{psfig,graphicx}

\def\gapprox{\lower.4ex\hbox{$\;\buildrel >\over{\scriptstyle\sim}\;$}}
\def\lapprox{\lower.4ex\hbox{$\;\buildrel <\over{\scriptstyle\sim}\;$}}
\def\bk{\mbox{\boldmath $k$}}
\def\bn{\mbox{\boldmath $n$}}
\def\bv{\mbox{\boldmath $v$}}
\def\be{\mbox{\boldmath $e$}}
\def\bB{\mbox{\boldmath $B$}}

\def\bnabla{\mbox{\boldmath $\nabla$}}

\title[Propagation of radio emission in pulsar magnetospheres]{The 
induced turbulence effect on propagation of radio emission in pulsar magnetospheres}

\author[Luo \& Melrose]
      {Qinghuan Luo and D. B. Melrose\\
        School of Physics, The University of Sydney, NSW 2006, Australia\\
}
                                                                                                            
\date{
          --- Received
         in original form April, 2006
        }
                                                                                                            
\pubyear{2006}
                                                                                                            
\begin{document}
                                                                                                            
\maketitle
                                                                                                            
\begin{abstract}
The effect of photon-beam-induced turbulence on propagation of 
radio emission in a pulsar magnetosphere is discussed. 
Beamed radio emission with a high brightness temperature can generate 
low-frequency plasma waves in the pulsar magnetosphere and these
waves scatter the radio beam. We consider this 
effect on propagation of radio emission both in the open field line region and 
in the closed field line region. The former is applicable to most cases of 
pulsar radio emission where the propagation is confined to the polar region;
it is shown that the induced process is not effective for radio emission
of moderately high brightness temperature but can have a severe effect on
giant pulses. For giant pulses not to be affected by this process, they must be
emitted very close to the light cylinder.
We show that the induced process is efficient in the closed field
line region, inhibiting propagation of the radio emission in this region. 
\end{abstract}

                                                                                                            
\section{Introduction}

It is generally thought that pulsar radio emission is produced in the 
open field line region well inside the light cylinder (LC) (the radius 
at which  the corotation speed equals $c$). Radio emission generated inside the 
LC must propagate over a substantial part of the magnetosphere before escaping to 
the interstellar medium. The intense radio emission is subject to various interactions 
with the intervening  plasmas within the pulsar magnetosphere. Two types of interactions 
have been considered: cyclotron absorption in which electrons (positrons) are in 
cyclotron resonance with the wave (Blandford \& Scharlemann 1976; Fussell, Luo, \& Melrose 2001), 
and induced Compton scattering in which the radio wave is scattered to a higher or 
lower frequency~\citep{bs76,p04b}.  Both processes
can affect or even disrupt the propagation of the beam. So far, no clear 
observational evidence for the absorption feature associated with the cyclotron 
resonance has been identified. However, the conventional polar cap theory, which 
assumes that the radio emission is generated in an outflowing electron-positron pair 
plasma, does predict a cyclotron resonance region where absorption should occur. 
Induced Compton scattering can be important in principle for any coherent emission with 
a sufficiently high intensity~\citep{wr78,sk92,cbr93,hm95}. 
While cyclotron absorption is important only in the 
region near the LC, induced scattering can occur throughout the magnetosphere.
Apart from these two processes, there is a third process that
involves induced wave-wave interactions and that can significantly affect 
propagation of the radio beam. A highly collimated beam of 
radio waves can generate low frequency waves through three-wave interactions
in the pulsar magnetospheric plasma in a way similar to wave generation in
a particle beam instability. These low frequency waves interact with the radio 
waves, leading to a decrease in its intensity and an increase in 
its angular spread. In some cases, such effects can completely disrupt 
the radio beam: The medium becomes opalescent to the radio waves. 

In this paper, we discuss the effect of induced three-wave interactions on 
propagation of radio emission in a pulsar magnetosphere. Induced
three-wave processes have been considered for pulsar winds \citep{lm94},
pulsar eclipse by the stellar wind of its companion~\citep{t-etal94,m94,lm95}
and radio emission from active galactic nuclei~\citep{k88,gk93,lb95}.
Apart from a discussion of Raman scattering (a particular case
of three-wave interactions, cf. Sec. 2.5) in a pulsar magnetosphere~\citep{gk93,l98}, we are aware 
of no detailed study of the three-wave effect which explicitly takes account of 
the beaming and broadband nature of the radio emission. To our knowledge, there is no
detailed study of propagation of radio emission in the closed field line region. Here
we consider the three wave effect both in the open field line region (OFLR) and in 
the closed field line region (CFLR). In most models, propagation of radio emission is 
confined to the OFLR where the relativistic electron positron pair plasma 
streams along magnetic field lines at a relativistic velocity. Because electrons and positrons
contribute to three-wave interactions equally but with opposite sign, 
three-wave interactions are significant weakened in a quasineutral pair plasma.
However, such processes can still be important for radio emission with very
brightness temperature provided that there is a modest excess of electrons 
or positrons and that the bulk Lorentz factor is modest. Since for young pulsars,
the conventional polar cap model generally predicts a plasma density too high
for propagation of radio waves, we assumes that the plasma is highly inhomogeneous
across the polar cap and that radio emission is generated
at a frequency near the plasma frequency (in the plasma's rest frame) in the underdense
region where the plasma density is relatively low (cf. Sec. 3.1).  
This assumption seems to be required for consistency if the 
radio emission is produced due to a beam instability \citep{mg99}. 

Propagation of radio emission in the CFLR is relevant at least for three specific 
cases: (1) the recently discovered double pulsar system PSR J0737-3039B
in which the radio emission from one pulsar interacts with the plasma in 
the CFLR of the other, (2) radio emission from outer magnetospheric regions
where the emission of the trailing component may propagate to the CFLR as a result of
the aberration effect, and (3) backward emission predicted by the oscillatory polar 
gap model~\citep{lmjl05} and by recent models for the peculiar `notch-like' feature 
of pulse profiles~\citep{d-etal05}. For case (1), eclipse of the radio emission from
one pulsar by the magnetosphere of the other has been observed. A possible
interpretation of the eclipse is due to processes in the CLFR
such as cyclotron or synchrotron absorption  or induced scattering. 
For case (2) the trailing component may have an absorption-like feature or may be
completely destroyed as the result of strong induced processes. 
For the third case, induced processes can severely constrain
visibility of the backward emission that propagates through the CFLR.

In Sec. 2, we outline a general formalism of induced three-wave interactions in 
the pulsar magnetosphere. We discuss the effect on propagation
of radio emission in the OFLR in Sec. 3 and in CFLR in Sec. 4.

\section{Induced three-wave interactions}

We consider three-wave processes due to passage of beamed radio 
emission (referred to as a photon beam or radio beam).
The processes involve two high frequency waves (unscattered 
and scattered radio waves) and a low frequency wave 
generated by the radio beam. The three waves satisfy the 
resonance conditions $\omega\leftrightarrow\omega'+\omega''$ 
and $\bk\leftrightarrow\bk'+\bk''$. The condition 
$\omega\sim\omega'\gg\omega''$ means that three wave scattering
is approximately `elastic'.

\subsection{Three-wave probability}

We model pulsar radio emission as a highly collimated photon 
beam, described by the photon occupation number $N(\bk)$,
where $\omega$ is the wave frequency and $\bk$ is the wave vector.
The photon occupation number is a Lorentz invariant, related to the
brightness temperature $T_b$ through $\hbar\omega N(\bk)=k_BT_b$,
where $k_B$ is the Boltzmann constant. 
Let $N_L(\bk'')$ be the occupation number of the low-frequency 
wave with frequency $\omega''$ and wave vector ${\bk}''$. 
Evolution of $N(\bk)$, $N(\bk')$ and $N_L(\bk'')$ are determined by
three kinetic equations which are completely determined by the three-wave probability
\citep{m86}:
$W(\bk,\bk'\bk'')(2\pi)^4\delta(\bk-\bk'-\bk'')\delta(\omega(\bk)-\omega'(\bk')-\omega''(\bk''))$,
where the delta functions are the three-wave resonant conditions  and 
\begin{equation}
W(\bk,\bk',\bk'')=\hbar R_L{|e_ie'_j{e''}^*_s\alpha_{ijs}|^2\over
\varepsilon^3_0\omega^2\omega''}.
\label{eq:W}
\end{equation} 
In (\ref{eq:W}), which is simply referred to as the three-wave probability,
$\alpha_{ijs}$ is the quadratic response tensor,
$R_L$ is the ratio of the electric to total energy of the low frequency wave, and 
the polarizations of three waves are ${\be}$, ${\be}'$ and ${\be}''$. 
In deriving the three-wave probability, we adopt the
cold plasma form for the quadratic response tensor, e.g.
Eq (A1) in~\cite{lm06}, and take the strong magnetic field limit.
One should note that the strong field limit may not be appropriate for the region near
the LC. Nonetheless, it allows us to simplify the quadratic response
tensor greatly. In this limit, one finds that the dominant component is the
projection to the magnetic field, denoted by $\alpha_\parallel$, given by
\begin{equation}
\alpha_\parallel\approx {\eta \varepsilon_0 e\omega^2_p\over2m_ec}\left(n_\parallel
+n'_\parallel-n''_\parallel\right),
\end{equation}
where $\omega\approx\omega'\gg\omega''$, $\omega_p$ is the plasma
frequency, $n_\parallel=k_\parallel c/\omega$, $n'_\parallel=k'_\parallel
c/\omega'$, $n''_\parallel=k''_\parallel c/\omega''$,
$\eta$ is the ratio of the difference to the sum of the positron and electron
number density, with  $\eta=-1$ corresponding to an electron gas. 
We use the subscript $\parallel$ to denote the parallel (to the magnetic field) 
component. The final expression for the three-wave probability is
\begin{equation}
W\approx
{\pi\over2}\eta^2 r_ec^2\,{\hbar\omega\over m_ec^2}\,\left({
\omega_p\over\omega}\right)^3
|\xi_p|^2,
\label{eq:W2}
\end{equation}
where $r_e=e^2/(4\pi\varepsilon_0m_ec^2)\approx2.8\times10^{-15}\,\rm m$ 
is the classical electron radius  and
$\xi_p=(\omega_p/\omega'')^{1/2}(n_\parallel+n'_\parallel-n''_\parallel)e_\parallel 
e'_\parallel e''_\parallel $. The expression (\ref{eq:W2}) 
is closely analogous to a similar expression for an unmagnetized plasma~\citep{m94}
where the low frequency waves are the Langmuir waves ($\omega''=\omega_p$). 
For a quasi-neutral, cold electron-positron plasma, one has $\eta=0$;
one then needs to retain higher order terms in an expansion in
$(\omega,\omega'')/\Omega_e\ll1$ to obtain a nonzero three-wave probability, which
is much smaller than (\ref{eq:W2})~\citep{lm97}. 

In the following, we consider two low frequency modes that may be
relevant to induced three-wave interactions: the Langmuir mode that propagates along the 
magnetic field, and the Alfv\'en mode (also called the low frequency $LO$ mode) 
which can propagate at any direction. Note that the three-wave process involving a 
Langmuir wave in the (coherent) fixed-phase formalism is also called stimulated Raman scattering
(cf. Sec. 2.5).  Landau damping imposes a lower limit to the wavelength of low frequency waves.
To determine the Landau damping one need consider separately an intrinsically 
relativistic plasma such as in the OFLR and a nonrelativistic plasma such as 
in the CFLR. Landau damping in an intrinsically relativistic plasma 
occurs preferentially for $n''\sim1$ since
the plasma is relativistic even in the plasma rest frame and the particle's
Cerenkov resonance occurs for the wave phase speed near the speed of light. 
In the nonrelativistic plasma, Landau damping is effective for electrostatic waves with
$n''\sim n''_{\rm max}=0.3c/\lambda_D\omega_{p}$ (where $\lambda_D$ is the Debye length)~\citep{b66},
leading to a limit on the maximum $n''_{\rm max}\gg1$. 
Since $|\xi_p|^2\propto {n''}^2$, three-wave interactions involves 
low frequency electrostatic waves should be important in the CFLR where the plasma
is expected to be nonrelativistic (cf. Sec. 4.1). In the OFLR the plasma is 
intrisically relativistic, Landau damping limits
the refractive index of the Langmuir waves to about 1, and both the Alfv\'en 
and electrostatic modes need to be considered.

\subsection{Scattering rate}

We consider the production rate of scattered photons as a measure of the 
effect of the three-wave process on the photon beam. This rate, which we also call
the scattering rate, depends on the level of the low frequency waves 
generated by the photon beam. We consider the case 
where the low frequency waves are saturated as the result of change in
the occupation number of the photon beam. The saturation
condition $dN_L/dt=0$ leads to a level of low frequency waves expressed
in terms of the photon occupation number. Thus the scattering rate can be
written as
\begin{equation}
\Gamma={d\ln N(\bk')\over dt}\approx
{\Delta\Omega k^2\over(2\pi)^2}\,{WN(k)\over c},
\label{eq:Gamma}
\end{equation}
where $\Delta\Omega$ is the angular spread of the photon beam. Our specific model
for an axially symmetric beam is
\begin{equation}
N(\bk)=N(k)b(\Theta),\quad\quad
b(\Theta)=\exp\left(-{\Theta^2\over2\Delta\Theta^2_0}\right),
\end{equation}
where $\Delta\Omega=2\pi\Delta\Theta^2_0$, $\Delta\Theta_0\ll1$ is the 
beam width and the polar angles $(\Theta,\Phi)$ are defined with respect to the beam axis.
Similarly, we use $(\Theta'',\Phi'')$ for the polar angles of ${\bk}''$. 
The polar angles of $\bk$ and ${\bk}''$ relative to the magnetic
field ($z$ axis) are given respectively by $(\theta,\phi)$ and $(\theta'',\phi'')$.
Since the angular spread of the beam is small, hereafter we do not distinguish between
the beam axis and $\bk$.  Substituting (\ref{eq:W2}) for (\ref{eq:Gamma}) yields
\begin{eqnarray}
\Gamma\approx {\eta^2r_e\omega^2_p\over 8\pi c}\,\Bigl({\Delta\Omega k_BT_b\over m_ec^2}
\Bigr)\,{\omega_p
\over\omega}\,\left|\xi_p\right|^2,
\label{eq:Gamma2}
\end{eqnarray}
where we express the incoming photon occupation number in terms of 
the brightness temperature: $\hbar\omega N(k)=k_BT_b$, and
$\xi_p$ depends on the specific low frequency mode involved and is derived in Sec. 2.3 and 2.4.

From (\ref{eq:Gamma2}) one may define the opacity for the photon beam propagating 
over a distance $\Delta s$:
\begin{eqnarray}
\tau&\approx&\Gamma{\Delta s\over c}\nonumber\\
&\approx&{\eta^2r_e\Delta s\omega^2_p\over 8\pi c^2}\,\Bigl({\Delta\Omega k_BT_b\over m_ec^2}
\Bigr)\,{\omega_p
\over\omega}\,\left|\xi_p\right|^2.
\label{eq:tau}
\end{eqnarray}
In (\ref{eq:tau}), $\Delta\Omega T_b$ can be estimated from the observed flux
density $S^{\rm obs}_\nu$:
\begin{eqnarray}
\Delta\Omega T_b&\approx&{c^2\over k_B\nu^2}\,\left(
{D_L\over \Delta s}\right)^2S^{\rm obs}_\nu\nonumber\\
&=&5\times10^{22}\,{\rm K}\,\left({\Delta s\over10^5\,{\rm m}}\right)^{-2}
\left({1\,{\rm GHz}\over\nu}\right)^2
\left({D_L\over 1\,{\rm kpc}}\right)^2\nonumber\\
&&\times\left({S^{\rm obs}_\nu\over
10\,{\rm mJy}}\right),
\label{eq:Tb}
\end{eqnarray}
where $\nu\equiv\omega/2\pi$ is the frequency in Hz, $\Delta s$ is the 
distance from the emitting region to the region concerned and $D_L$ is the pulsar distance.
If one assumes the size of the emission region is $\delta l$, one
has $\Delta\Omega\sim (\delta l/ \Delta s)^2$. The brightness
temperature is estimated to be $T_b\sim 5\times10^{28}\,\rm K$ for
$\delta l=10^2\,\rm m$.

\subsection{Back scattering}

Consider first the case where the low frequency waves are Langmuir waves 
propagating along the magnetic field at a low phase speed 
$\omega''/k''\ll c$ (or $n''\gg1$).  Three-wave interactions favor
the back scattering regime where the incoming and scattered photons propagate in 
opposite directions. From the three-wave resonance conditions one has
$k_\parallel=k'_\parallel+k''$ (parallel to the magnetic field)
and $\bk_\perp=\bk'_\perp$ (perpendicular to the magnetic field),
which gives rise to $\cos\theta\approx\cos\theta'+n''(\omega_p/\omega)$, where
$n=kc/\omega\approx n'=k'c/\omega'\approx1$ and $\omega''=\omega_p$. Because of $k\approx k'$, one must
have $\theta'\sim \pi-\theta$, corresponding to the incoming and scattered 
photons propagating in opposite directions. Hence for Langmuir waves, one derives
\begin{equation}
|\xi_p|^2\approx {n''}^2\sin^4\theta\approx
4\left({\omega\over\omega_p}\right)^2\cos^2\theta\sin^4\theta,
\label{eq:xi}
\end{equation}
provided that $2\cos\theta(\omega/\omega_p)<n''_{\rm max}$, where $n''_{\rm max}$ is determined
by the Landau damping. The angular range for back scattering to occur is
$\cos\theta\leq {\rm min}\{1,k''_{\rm max}/2k\}$ with $k''_{\rm max}=n''_{\rm max}\omega_p/c$.
When the photon beam is parallel to the magnetic field,
one has $|\xi_p|^2\sim (\omega/\omega_p)^2(\Delta\Omega/\pi)^2$, which implies that
only photons off the beam axis with $\theta\neq0$ can interact with the Langmuir waves.
In a nonrelativistic plasma, one has $n''_{\rm max}\gg1$
and hence the back scattering can be very efficient.

\subsection{Diffusion of the photon beam}

In the regime $k''\ll k\sim k'$, the three-wave interactions resemble 
particle-wave interactions, with high frequency waves (the radio emission) acting
like `particles' that can emit or absorb low frequency waves \citep{m94,lm95,lm97}. 
This regime is also called the small angle scattering because the angle between 
the incoming and scattered photons is small, given by $\theta_{kk'}\approx k''/k\ll1$.
The three-wave resonance conditions reduce to the Cerenkov resonance condition
between `particles' (the high frequency wave) with a velocity ${\bv}_g=
\partial\omega(\bk)/\partial \bk$ and a wave with $(\omega'',{\bk}'')$.
Specifically, one may define a Cerenkov angle $\chi_0=\arccos(\omega''/k''c)$, 
corresponding to a conical surface with respect to ${\bv}_g$, which
defines the Cerenkov resonance condition. The condition can only
be satisfied for a subluminal wave $n''=k''c/\omega''>1$, i.e. 
its phase speed must be less than $c$. 
We are interested in negative absorption (or growth) of the low frequency
waves, which occurs for $\Theta''\sim \chi_0-\Delta\Theta_0$, at a rate determined by
the three-wave probability (\ref{eq:W}) or (\ref{eq:W2})~\citep{lm97}. 

The instability of low frequency waves leads to diffusion of photons
in the $\bk$-space, determined by the diffusion coefficient $D_{ij}
\propto {k''}^2k''_ik''_jWN_L/c$, where the three-wave probability
$W$ is given by (\ref{eq:W}). The growth of low frequency waves is entirely due to the 
anisotropic angular beaming of the photons. A decrease in the beaming, over the 
perpendicular diffusion time $t_\perp=k^2_\perp/D=\Delta\Omega k^2/D$,
where $D=\sum_iD_{ii}$, is especially important as it can ultimately suppress the growth.
Assuming that the low frequency waves are saturated due to perpendicular diffusion,
one can derive the total diffusion time $t_d=k^2/D=1/\Gamma$, which describes the 
effect of the instability on the beam. One finds $\Gamma$ is the
same as that given by (\ref{eq:Gamma}) or (\ref{eq:Gamma2})~\citep{lm95} and
the diffusion effect can be estimated by evaluating the opacity (\ref{eq:tau}). 
For Langmuir waves, one finds
\begin{eqnarray}
|\xi_p|^2&\approx& \left({2\over n''}\cos\theta-{\omega_p\over\omega}\sin\theta
-1\right)^2\nonumber\\
&&\times\left(1+{n''\omega_p\over\omega}\cot\theta\right)^2{n''}^2\sin^4\theta,
\label{eq:xi2}
\end{eqnarray}
where $1<n''\ll \omega/\omega_p$. Similarly for Alfv\'en waves one has
\begin{equation}
|\xi_p|^2\approx \left({\omega''\over\omega_p}\right)^3(2\cos\theta-1)^2
\sin^4\theta\tan^2\theta'',
\label{eq:xi3}
\end{equation}
where the parallel component of the polarization is $|e''_\parallel |\approx
(\omega''/\omega_p)^2n''\sin\theta''\ll1$, $\theta''<1$ and 
$n''_\parallel\approx1$~\citep{ab86,mg99}.

\subsection{Induced Compton scattering}

The relation between the induced three-wave interaction discussed here and induced (or stimulated)
Compton scattering requires comment. (We use `induced' and `stimulated' interchangeably here.) Both
processes need to be considered in discussing the propagation of radiation with very high brightness
temperature, and we regard them as independent of each other. In our view, the difference between
these two processes has become somewhat obscured by discussions of stimulated Raman scattering.
Formally, Raman scattering is associated with  particles (atoms or molecules) that have an intrinsic
transition frequency, $\omega_R$ say, associated with them. Spontaneous Raman scattering of incident
radiation with frequency $\omega$ produces scattered radiation at the Stokes frequency,
$\omega-\omega_R$, and at the anti-Stokes frequency, $\omega+\omega_R$. Stimulated Raman scattering
is analogous to absorption of the beat between the incident and scattered radiation, and occurs at a
rate that is proportional to the product of the intensities at the incident and the scattered
frequencies. This leads to exponential growth at the Stokes frequency, and an associated exponential
decay at  the incident frequency, so that radiation is transferred from higher to lower frequency at
an exponentially increasing rate. (If the scattering particles have an inverted energy distribution,
negative absorption or maser action results in an exponentially increasing transfer from the incident
frequency to the anti-Stokes frequency.) 

Induced Compton scattering is analogous to induced Raman
scattering, except that there is a continuum of frequencies for the scattered radiation. For
isotropic radiation with a peak in the intensity at some frequency, induced Compton scattering tends
to move this peak to lower frequencies; if there are two distinct beams of radiation, induced Compton
scattering tends to transfer radiation from the beam at the higher to the beam at the lower
frequency. The direction of transfer is reversed if the scattering particles have an inverted energy
population, and this has been suggested as a mechanism for generating giant pulses~\citep{p04a}.
The induced three-wave interaction discussed here is essentially the same process
described as `stimulated Raman scattering' by Gangadhara and Krishan  (1993) and as `induced Raman
scattering' by Lyutikov (1998) except that in our case the low frequency wave is not
restricted to plasma waves.  The important distinction, as noted by Lyutikov (1998), is that in
the induced three-wave interaction, the Raman frequency, $\omega_R$, is replaced by the frequency of
the low frequency waves, which is the plasma frequency, $\omega_p$, in cases of interest. However,
this analogy is only kinematic, and the dynamics of the two processes are quite different; for
example, they are described by different kinetic equations. In the induced three-wave interaction the
plasma may be regarded as passive, determining the wave dispersion and the nonlinearity that allows
the three-wave coupling. The kinetic equation involves on the wave distributions: the low frequency
waves grow and result in the scattering of the high-frequency waves. In contrast, in induced Compton
scattering, the plasma particles actively scatter the high-frequency waves, the process depends
explicitly on the distribution function for the particles, and no low-frequency wave is involved.

\section{Propagation in the OFLR}

The formalism derived in the preceding section can be applied to the case of
propagation of radio emission in the OFLR where the plasma has a bulk velocity
$v=\beta c$ and a Lorentz factor $\gamma=1/(1-\beta^2)^{1/2}$.
A convenient approach is to treat the formalism in Sec. 2 in the
plasma rest frame and then derive the relevant quantities in the 
pulsar or observer's frame by a Lorentz transformation. 
Note that some caution is needed in defining the bulk rest 
frame. Since the plasma streams along curved magnetic field lines,
there is no single bulk frame. However, one can still construct a consecutive 
series of local bulk frames along the curved field lines, an approach 
similar to that used in ray tracing in a relativistically flowing plasma~\citep{fl04}. 
We assume that the photon beam is axially symmetric in the plasma rest frame. 
The photon beam shape varies from frame to frame. Such complication is not considered here, 
though it should be included in a full radiative transfer calculation. 
Specifically, the relevant propagation angle is transformed according to
\begin{equation}
\tilde{n}\sin\tilde{\theta}={n\sin\theta\over\gamma(1-n\beta\cos\theta)},
\quad
\tilde{n}\cos\tilde{\theta}={n\cos\theta-\beta\over1-n\beta\cos\theta}.
\label{eq:LT}
\end{equation}
All tilde quantities correspond to the rest frame.
We ignore the pulsar's rotation in the Lorentz transformation.  The brightness 
temperature, frequency, differential solid angle and plasma frequency
are transformed according to $\tilde{T}_b=\gamma(1-n\beta\cos\theta)T_b$, 
$\tilde{\omega}=\gamma(1-n\beta\cos\theta)\omega$,
$\Delta\tilde{\Omega}=\Delta\Omega/\gamma^2(1-n\beta\cos\theta)^2$,
and $\tilde{\omega}_p=\omega_p/\gamma^{1/2}$,
where the refractive index is $n\approx1$. 
Since the time derivative in (\ref{eq:Gamma}) is interpreted as the total
time derivative, i.e. $d/dt=\partial/\partial t+c\bn\cdot\bnabla$ with $\bn=\bk/k$,
one has $\tilde{\Gamma}=\Gamma/\gamma(1-\beta\cos\theta)$~\citep{p73,hm95,ss98}.
Upon making the Lorentz transformation one obtains the scattering rate 
in the observer's frame:
\begin{eqnarray}
\Gamma&\approx& {r_e\omega^2_{pe}\over 8\pi (2M\gamma)^{1/2}c}
\,\Bigl({\Delta\Omega k_BT_b\over m_ec^2}
\Bigr)\,{\omega_{pe}
\over\omega}{2\left|\tilde{\xi}_p\right|^2\over1+\theta^2\gamma^2},
\label{eq:Gamma3}
\end{eqnarray} 
where $M$ is the pair multiplicity. In (\ref{eq:Gamma3}), 
we use $\eta=1/2M$ and rewrite the plasma frequency as $\omega_p=(2M)^{1/2}\omega_{pe}$, where $\omega_{pe}$
is the plasma frequency corresponding to the number density of the excess 
electrons or positrons. 

\subsection{Plasma density}

The scattering rate is proportional to the local net charge density:
a higher net charge density implies a faster scattering rate.  
This is due to the electrons and positrons contributing with opposite 
sign to the three-wave coupling coefficient. For the conventional polar 
cap model the number density of the excess charge is the Goldreich-Julian (GJ) density, 
given by $n_{GJ}\approx7\times10^{17}\,{\rm m}^{-3} (P/0.1\,{\rm s})^{-1}(B/10^{8}\,
{\rm T})(R_0/r)^3$~\citep{gj69}, where $R_0=10^4\,\rm m$ is
the star's radius. Seemingly insurmountable difficulties arise if
one assumes the most widely discussed emission mechanism (a beam instability)
in a plasma with the GJ density times a multiplicity $M\geq1$.
The lowest frequency consistent with theory is 
$\omega_c\sim 2(\gamma/\langle\gamma'\rangle)^{1/2}\omega_p$
~\citep{mg99}, where $\langle\gamma'\rangle$ is the Lorentz factor's mean spread in the 
plasma rest frame, about 1-10~\citep{ae02}. In the following we assume 
$\langle\gamma'\rangle=4$ implying $\omega_c=\gamma^{1/2}\omega_p$. 
For example, the plasma frequency predicted for the Crab 
pulsar is $\omega_p\approx (2\times10^{12}\,{\rm s}^{-1}\,)(R_0/r)^{3/2}(M/100)^{1/2}$, 
corresponding to $\omega_c\sim \gamma^{1/2}\omega_p\approx
(2\times10^{13}\,{\rm s}^{-1})(R_0/r)^{3/2}(M/100)^{1/2}$ for $\gamma=100$, which is 
too high compared with the observed frequency. 
(This plasma density problem was discussed by \cite{ketal98}.) 
One practical solution to this problem is that the plasma is highly inhomogeneous forming 
underdense regions where radio emission is produced and propagates. The 
low density region where the density is actually much lower than $2Mn_{GJ}$
can form as a result of nonuniform extraction of primary particles 
from the polar cap or nonuniform pair cascades~\citep{lmjl05}. 
In the former case, the primary particle density is nonuniform 
across the polar cap forming localized outflowing fluxes along field lines separated 
by low density regions where the net charge density is much less than $n_{GJ}$ 
(though on average the mean flux density from the polar cap is still $n_{GJ}c$).
Both the multiplicity $M$ and the net charge density can be low in the underdense region. 

In the following we consider the case where the radio 
wave is emitted in the low density regions at a frequency near the local plasma frequency.
Specifically we assume (1) the radio frequency is $\omega\sim\omega_{p*}\gamma^{1/2}$,
where $\omega_{p*}\leq(e^2\bar{M}n_{GJ}/\varepsilon_0m_e)^{1/2}(R_0/r_*)^{3/2}$ is the 
local plasma frequency at the emission radius $r_*$ and $\bar{M}$ is the mean multiplicity, and (2)
$M\leq\bar{M}$ is a constant along the propagation path. Assumption (1) is applicable if the coherent 
radio emission is due to wave-particle interactions \citep{mg99}. Although assumption (2) may not be realistic, 
particularly in the oscillatory case where both the multiplicity $M$ and the net charge density
can be nonstationary \citep{lmjl05}, it is a reasonable approximation provided that 
the size the scattering region
($r_*/\gamma$) is much larger than the length scale (corresponding to the polar gap length) 
of plasma inhomogeneities along the propogation path. These two assumptions allow us 
to estimate (\ref{eq:Gamma3}) without going into much details on a particular polar 
cap model. We consider the following three propagation regimes separately: two 
quasi-parallel regimes, including the forward propagation $\tilde{\theta}\ll1$ 
and backward propagation $\tilde{\theta}\sim \pi$, and quasi-perpendicular 
regime $\tilde{\theta}\sim \pi/2$.

\subsection{Quasi-parallel propagation}

In the forward propagation regime $\tilde{\theta}\ll 1$,  ${\bk}$ is nearly parallel
to the magnetic  field. Assuming the low frequency plasma
wave propagates along the magnetic  field, one has  $\tilde{\theta}''=0$ and
$\tilde{n}''\sim 1$. We assume that the photon beam is in the $LO$ mode with 
the polarization in the $\bB$-$\bk$ plane, because only this high frequency mode 
can be produced through wave-particle interactions~\citep{mg99}. 
If one assumes that the radio emission is generated near the frequency 
$\omega\sim\omega_{p*}\gamma^{1/2}$ at the emission radius $r_*$, the plasma 
frequency in the rest frame is comparable with the photon frequency. Thus 
the only relevant low frequency waves are Alfv\'en waves.

For the Alfv\'en mode, one can show that three-wave interactions due to the 
projection of the polarizations of all three waves along the magnetic field are 
small. From (\ref{eq:xi3}), which applies in the rest frame,
one obtains $|\tilde{\xi}_p|^2 \sim
(\tilde{\omega}''/\tilde{\omega}_p)^3\tilde{\theta}^4\ll1$, where
$\tilde{e}_\parallel\sim \tilde{e}'_\parallel\sim \tilde{\theta}$.
For $\tilde{\omega}''/\omega_p=0.2$ and $\tilde{\theta}=0.1$, one 
has $|\tilde{\xi}_p|^2\approx 10^{-6}$. 

\begin{figure}
\psfig{file=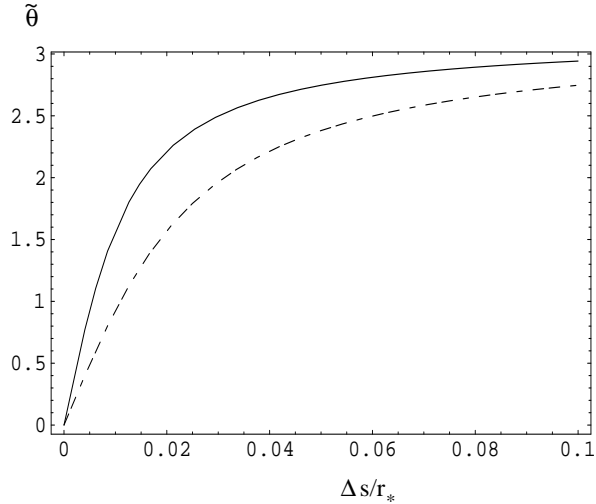,width=8cm}
\caption{Propagation angle in the rest frame vs propagation distance $\Delta s/r_*$. 
The propagation angle $\tilde{\theta}$, which is assumed to start from zero (at $\Delta s=0$), 
rapidly increases to $\pi$ after a distance $\Delta s=r-r_*>r_*/\gamma$. A dipole 
magnetic field is assumed. The solid and dashed lines correspond to $\gamma=100$ and 50,
respectly.}
\label{fig:angle}
\end{figure}

In most part of its propagation path the radio emission propagates backward in the 
rest frame, as indicated in figure~\ref{fig:angle}, and in 
this regime the low frequency approximation is justified. From (\ref{eq:LT}), 
one has $\cos\tilde{\theta}\approx (1-\theta^2\gamma^2)/(1+
\theta^2\gamma^2)\approx -1$ for $\theta\gg1/\gamma$, i.e. the emission 
propagates backward ($\tilde{\theta}\sim \pi$) in the rest frame.
In the rest frame the frequency is given by 
$\tilde{\omega}\approx \theta^2\gamma\omega/2$. Since $\omega\sim \omega_{p*}\gamma^{1/2}$,
one has $\tilde{\omega}/\tilde{\omega}_p \sim\theta^2\gamma^2\gg1$ or 
$\tilde{\omega}\gg \tilde{\omega}_p$, 
where $\tilde{\omega}_p\sim \tilde{\omega}_{p*}$ for
$\Delta s/r_*=(r-r_*)/r_*<1$.
Thus for backward propagation, the Langmuir waves can be
treated as low frequency waves ($\tilde{\omega}\gg \tilde{\omega}_p$) and
back scattering (cf Sec. 2.3) is allowed kinematically but 
is inhibited by the Landau damping. For backward propagation, the
$LO$ mode is elliptically polarized, with only a small component along the
magnetic field $|e_\parallel |\approx |e'_\parallel |\approx 2/\theta\gamma\ll1$.
From the three-wave resonance conditions one finds
$n''\sim 2\tilde{\omega}/\tilde{\omega}_{p}\sim \theta^2\gamma^2\gg1$.
However, Landau damping limits the refractive index to $\tilde{n}''\sim 1$.

\begin{figure}
\psfig{file=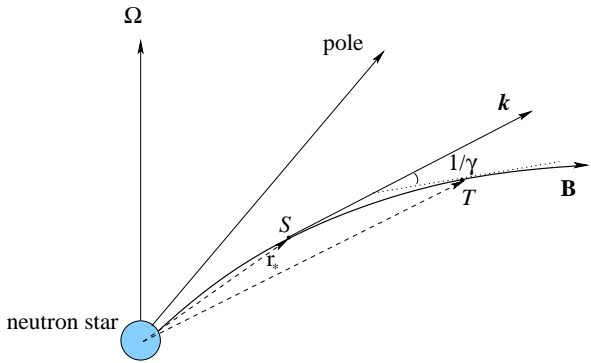,width=8cm}
\caption{Propagation of the photon beam emitted at $r_*$. The propagation 
enters the quasi-perpendicular regime at $T$ where the propagation angle 
in the pulsar frame is $\theta\sim 1/\gamma$. The distance from the 
source ($S$) (at $r_*$) to the quasi-perpendicular regime ($T$) 
is approximately $r_*/\gamma$.
}
\label{fig:qtr}
\end{figure}

For backward propagation, the three-wave interaction can proceed in the regime 
$k''\ll k\sim k'$ where both Langmuir and Alfv\'en waves are relevant. 
From (\ref{eq:xi2}) one obtains 
$2|\tilde{\xi}_p|^2/(1+\theta^2\gamma^2)\approx 2/\theta^6\gamma^6\ll1$ for the Langmuir waves with
$\tilde{n}''\approx n''\sim 1$ and from (\ref{eq:xi3}) one has $2|\tilde{\xi}_p|^2/
(1+\theta^2\gamma^2)\approx2
(\tilde{\omega}''/\tilde{\omega}_p)^3(\tilde{\theta}'')^2/\theta^6\gamma^6$ for the Alfv\'en mode. 
Thus, Alfv\'en waves are less efficient than Langmuir waves.

\subsection{Quasi-perpendicular regime}

The transition between forward and backward propagation, $\tilde{\theta}=\pi/2$, 
in the rest frame corresponds to $\theta\sim1/\gamma$ in the pulsar frame. 
A photon beam that is emitted initially along the magnetic field 
rapidly acquires an angle as the magnetic field lines bend away (cf.
figure~\ref{fig:angle}). The propagation enters the quasi-perpendicular
regime ($\tilde{\theta}\sim \pi/2$) after the distance $\Delta s\sim r_*/\gamma$.
Similarly to forward propagation, the plasma frequency is comparable
with the radio frequency in the rest frame. So, the plasma waves cannot be treated as low
frequency waves. The relevant low frequency waves are the Alfv\'en waves.
From (\ref{eq:xi3}) we have $|\tilde{\xi}_p|^2\approx (\tilde{\omega}''/\tilde{\omega}_p)^3
\tan^2\tilde{\theta}''$. Although the process favors a higher $\omega''$, the Alfv\'en mode is
heavily damped near the plasma frequency. It can freely propagate in the range $\tilde{\omega}\ll
\tilde{\omega}_p$. For example, one may choose $\tilde{\theta}''=0.5$ and 
$\tilde{\omega}''/\tilde{\omega}_p=0.2$. This gives $|\tilde{\xi}_p|^2\approx 10^{-3}$.

To calculate (\ref{eq:Gamma3}), we determine the plasma frequency in the propagation
path from $\omega_{pe}=\omega_p/(2M)^{1/2}=[\omega/(2M\gamma)^{1/2}](r_*/r)^{3/2}$ with
$r\sim r_*$ in the quasi-perpendicular region. Note that for young pulsars such as the Crab 
pulsar the local plasma frequency determined in this way is much lower than the 
one derived from the density $\bar{M}n_{GJ}(R_0/r_*)^3$ (cf. Sec 3.1). 
The opacity $\tau=\Gamma\Delta s/c=\Gamma r_*/\gamma c$ is estimated as
\begin{eqnarray}
\tau&\approx&0.012\left({S^{\rm obs}_\nu\over10\,{\rm mJy}}\right)\left({
D_L\over1\,{\rm kpc}}\right)^2\left({M\over 10^2}\right)^{-2}
\left({\gamma\over10^2}\right)^{-1} \nonumber\\
&&\times\left({r_*\over10^5\,{\rm m}}\right)^{-1}
|\tilde{\xi}_p|^2.
\label{eq:tau2}
\end{eqnarray}
where (\ref{eq:Tb}) with $\Delta s=r_*/\gamma$ is used.
It should be noted that if the emission is from a region spread over $r_*/\gamma$,
the opacity estimated from (\ref{eq:tau2}) is reduced.
If one assumes the nominal value $|\tilde{\xi}_p|^2=10^{-3}$ for typical 
pulsar parameters, Eq. (\ref{eq:tau2}) shows that the induced diffusion 
is not important for radio emission with a moderately high brightness temperature. 
However, it can be effective for young pulsars or pulsars that emit giant pulses.

\subsection{Giant pulses}

For giant pulses the opacity (\ref{eq:tau2}) can well exceed unity and as a result,
it imposes a severe constraint on their propagation.  Giant pulses have been observed 
from a few pulsars with the two best examples being
the Crab pulsar~\citep{letal95,hetal03}, a fast rotating young pulsar and
PSR 1937+21~\citep{kt00}, the second fastest millisecond pulsar. The distribution of 
pulse energy (the flux density times the typical pulse width) is power-law, in contrast to
normal radio pulses that have a log-normal distribution~\citep{rj01}. This 
implies that the emission mechanism for giant pulses is distinct from that for
normal radio pulses. Although the emission mechanism for giant pulses is not well
understood, recent study of giant pulses in relation to the corresponding 
high energy emission suggests that the emission region may be in or near the 
outer magnetospheric region where the high energy emission originates~\citep{rj01}. 
Here we show that the opacity constraint (\ref{eq:tau2}) excludes the possibility of
giant pulses originating from the inner magnetosphere and deduce that
they must come from the outer magnetosphere, near the LC.

The opacity (\ref{eq:tau2}) decreases with increasing emission radius. This feature 
can be understood as follows. The efficiency of induced three-wave processes
depends on the photon energy density. For an emitting source of size $\delta l$,
the photon energy density in the quasi-perpendicular region (at a distance $r_*/\gamma$ from the
source) is proportional to $(\delta l/r_*)^2k_BT_b$. Observations of nanosecond giant pulses
from the Crab suggest a source size of order a meter \citep{hetal03}. 
This density decreases with increasing $r_*$, leading to a weakening of three-wave 
interactions. For giant pulses from the Crab pulsar, one has an average flux density
$S^{\rm obs}_\nu\sim 10^3\,\rm Jy$. We have $\tau\gg1$ unless a large emission 
radius $r_*$ is assumed. For giant pulses not to be 
strongly affected by the induced process, corresponding to the condition $\tau<1$,
the emission radius must satisfy the condition
\begin{eqnarray}
r_*&>& (5\times10^5\,{\rm m})\,
\left({S^{\rm obs}_\nu\over10^3\,{\rm Jy}}\right)\left({
D_L\over2\,{\rm kpc}}\right)^2\left({M\over 10^2}\right)^{-2}\nonumber\\
&&\times\left({\gamma\over 10^2}\right)^{-1},
\label{eq:r*}
\end{eqnarray}
where $|\tilde{\xi}_p|^2=10^{-3}$.
Although the pair multiplicity is not well constrained, (\ref{eq:r*}) still gives
a strong constraint on the emission radius for a reasonable
parameter range. A recent numerical simulation
suggests that the conversion efficiency of the primary particle energy 
to pairs is well below unity~\citep{ae02}, corresponding to $\bar{M}\gamma\ll\gamma_b/2$, where 
$\gamma_b$ is the primary particle's Lorentz factor. 
For young pulsars, resonant inverse Compton scattering is the dominant emission
process for pair-producing photons and it is reasonable to assume $\gamma_b\sim 10^5$.
This gives an upper limit $\bar{M}\gamma\ll5\times10^4$. Applying (\ref{eq:r*}) to the Crab pulsar, 
assuming a distance $D_L=2\,\rm kpc$, $M<\bar{M}=10^2$ and $\gamma=10^2$
one finds $r_*>5\times10^5\,{\rm m}$, about $0.3$ of the LC radius
$R_{LC}\approx 1.6\times10^6\,\rm m$. One concludes that for giant pulses to propagate 
without being dispersed by induced three-wave interactions, 
they must be emitted very close to the LC radius. 
For the millisecond pulsar PSR 1937+21, one expects $M$ to be small and $\gamma$
to be large as pair production in the millisecond pulsar's magnetic field is
much less efficient. Assuming that the primary particle's energy is limited by
curvature radiation corresponding to $\gamma_b\sim10^7$, one has the upper bound 
$M\gamma<5\times10^6$. For example, for $D_L=3.6\,\rm kpc$, $M=10$, $\gamma=10^5$ and 
$S^{\rm obs}_\nu=300\,\rm Jy$, one has $r_*> 3\times10^4\,\rm m$, 
about half the LC radius.

\section{Propagation in the CFLR}

The formalism discussed in Sec. 2 can be directly applied to the CFLR where
the plasma consists of single-sign charged particles with the GJ density $n_{GJ}$ and is
stationary with a zero bulk velocity. All the relevant quantities discussed
in Sec. 2 are now referred to the pulsar frame.

\subsection{Plasmas in the CFLR}

In contrast to the polar region, the plasma in the CFLR is plausibly stationary in
the corotating frame.
Since particles are in the ground Landau state they do not have a perpendicular 
momentum, so there is no magnetic mirror effect that can trap a dense plasma. This 
feature is very different from the usual
planetary magnetosphere in which a plasma is trapped due to this effect.
Thus, in the absence of continuous injection of pairs the plasma density 
in this region is controlled by the pulsar's rotation in such
way that the charge density is $en_{GJ}$~\citep{gj69}.
The plasma in this region consists of particles with a single sign of 
charge (either electrons or positron or possibly protons), which can be
treated as an electron gas in one dimension (along the field lines). 
Assuming the magnetic field as a function of radius $r$:
$B\propto (R_0/r)^\delta$, where $\delta=3$ for a magnetic dipole, one may express
the density as $n_e=n_{\rm GJ}(R_0/r)^\delta$.
Using (\ref{eq:tau}) and (\ref{eq:Gamma2}), one obtains 
\begin{eqnarray}
\tau&\approx& 4.6\left({\Delta\Omega T_b\over3\times10^{22}\,{\rm K}}\right)
\left({\nu\over1\,{\rm GHz}}\right)^{-1}
\left({P\over50\,{\rm ms}}\right)^{-5}\nonumber\\
&&\times\left({B\over10^8\,{\rm T}}\right)^{3/2}
\left({0.2R_{LC}\over r}\right)^{9/2}\left({\Delta s\over 0.2R_{LC}}\right)|\xi_p|^2,
\label{eq:tau3}
\end{eqnarray}
where $\xi_p$ is given by (\ref{eq:xi}) for back scattering and (\ref{eq:xi2})
for the quasilinear diffusion approximation.
Notice the difference in the radial distance dependence in (\ref{eq:tau3})
compared to (\ref{eq:tau2}). This difference is due to the GJ density 
varying with radial distance proportional to the magnetic field, where
we ignore a cosine factor. The CFLR region is separated into
a positive charge region and a negative charge region, with the 
boundary between them called the null surface.  One may define a radius 
$r_{cr}$ where $\tau=1$. The opaque region corresponds to $r\leq r_{cr}$, where 
induced three-wave processes are important. This critical radius is derived as
\begin{eqnarray}
{r_{cr}\over R_{LC}}&\approx& 0.3
\left({\Delta\Omega T_b\over3\times10^{22}\,{\rm K}}\right)^{2/7}
\left({\nu\over 1\,{\rm GHz}}\right)^{-2/7}\nonumber\\
&&\times\left({P\over 50\,{\rm ms}}\right)^{-10/7}
\left({B\over 10^8\,{\rm T}}\right)^{3/7}|\xi_p|^{4/7}.
\label{eq:rcr}
\end{eqnarray} 
We are interested in the region sufficently far from the star's surface
such that the condition $\omega\gg\omega_{pe}$ is satisfied. We discuss 
only the case where the low frequency waves are electrostatic.

\subsection{Eclipse in the binary pulsar J0737-3039B}

We first apply (\ref{eq:tau3}) to the double pulsar system J0737-3039B, in which
radio emission from one pulsar is periodically eclipsed by the other. 
It is of interest to consider if induced scattering can provide 
an effective mechanism for such eclipse.
The system consists of pulsar A with the period 22.7 ms and  magnetic field
$6.3\times10^5\,\rm T$, and pulsar B with the period 2.8 s and magnetic field
$1.6\times10^{8}\,\rm T$. The two pulsars are in a nearly circular orbit (with 
the eccentricity $0.088$), with the orbital plane seen nearly edge-on, at an inclination angle $\sim
87^\circ$ to the line of the sight~\citep{letal04}.
Observations indicate that the nearest impact distance of the photon beam
from pulsar A to pulsar B is about $0.1R_{LC}-0.2R_{LC}$ (of pulsar B).
The plasma frequency at $r=0.1R_{LC}$ is about $\omega_{pe}\approx 2\times10^5\,{\rm s}^{-1}$.
Cyclotron absorption is probably unimportant because the plasma density is too low~\citep{lt05}.

Consider interaction of a photon beam from pulsar A with the magnetospheric plasma
of pulsar B. The brightness temperature of the radio emission from pulsar A is 
estimated to be $T_b\sim (4.8\times 10^{16}\,\rm K)/\Delta\Omega$ at $1.36\,\rm GHz$ at 
$r=0.1R_{LC}\approx 1.4\times10^7\,\rm m$. To estimate the opacity, one may take 
$r\sim\Delta s=0.1R_{LC}$. 
Landau damping limits the Langmuir waves to $n''<n''_{\rm max}\approx
2\times10^3(T_e/10^6\,{\rm K})^{-1/2}(n_e/2.5\times10^7\,{\rm
m}^{-3})^{1/2}$, where $T_e$ is the plasma temperature.
For $P=2.8\,\rm s$, and $n''=2\times10^3$, one estimates 
$\tau\approx 10^{-5}$. The possibility of injection of relativistic pairs in the 
magnetosphere of pulsar B was recently discussed by \cite{rg05}, also \cite{lt05}.
In Rafikov \& Goldreich's model, trapping of a dense pair plasma is possible due to 
particles' nonzero pitch angles that are acquired as a result of synchrotron 
absorption of radio emission from pulsar A. Since the opacity is proportional to the
net charge density and is unchanged by the addition of pairs. Thus, we conclude that the 
induced three-wave process is not important for this binary system.

\subsection{Asymmetry in double-peak profiles}

Induced three-wave processes may be relevant in the outer magnetosphere for radiation 
originating near the last open field lines. Waves can be swept into the CFLR and
this effect is especially important for
a class of pulsars with double peaks with a wide separation of nearly $180^\circ$. 
In the single pole model, the main (the leading main) and inter (the weak trailing) 
pulses originate from emission from the opposite
sides of a wide cone at rather high altitudes. Two components are often asymmetric in
intensity with one (normally the leading component) being stronger than the other. 
One plausible interpretation of
such asymmetry is that waves from the trailing component are partially
absorbed due to cyclotron absorption~\citep{fl04}. 
Since the trailing component can be swept into the CFLR, 
induced three-wave process is a possible alternative mechanism
for its suppression.

\subsubsection{Radio beam crossing the CFLR}

\begin{figure}
\psfig{file=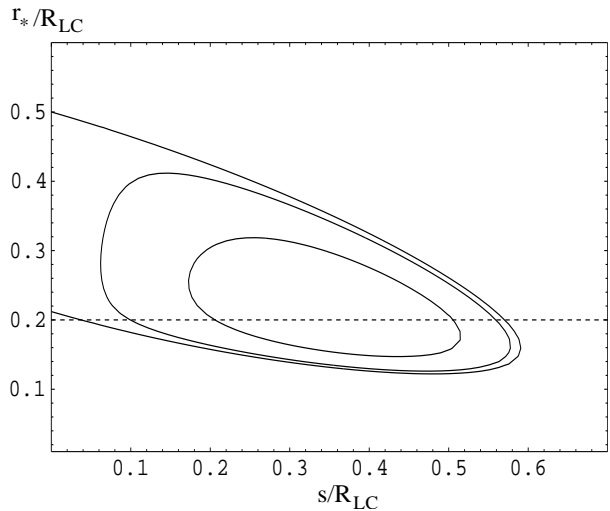,width=8cm}
\caption{The condition for a wave propagating into the closed field lines.
The contours (from innermost) correspond to closed field lines of $\zeta=0.995$, 
$\zeta=0.999$, and $\zeta=1$. The emission region is assumed to be
on the last open field lines ($\zeta=1$). The dashed line represents the 
propagation path of a wave emitted at $r_*=0.2R_{LC}$. The wave enters the 
CLFR at $s=0.04R_{LC}$ and exits at $s=0.57R_{LC}$. 
}
\label{fig:gc}
\end{figure}

To derive the condition for the radio emission to propagate
across the CFLR we assume that the photon trajectory
is straight in the observer's frame. In the co-rotation frame, the 
photon trajectory appears as a spiral curve. We consider an orthogonal rotator with the line 
of sight crossing the magnetic axis in the equatorial plane and emission from one pole 
only. In the corotating frame, one assumes a static dipole with 
field lines described by $r=\zeta^2R_{LC}\sin^2\Psi$,
where $\zeta=\sin\Psi_d/\sin\Psi_B$, $\Psi_d$ is the half opening angle of the polar cap,
and $\Psi_B$ is the colatitude of the field line foot point on 
the star's surface. The last open field lines correspond to $\zeta=1$.
Assume that waves are emitted at the radial distance $r_*$ on the last
open field field lines. After propagating a distance $s$ the radial
distance is approximately  $r_*+s$. If the photon beam propagates along 
a straight line given by $(r_*+s,\sigma)$ in the observer's frame, where $\pi/2-\sigma$ is the polar angle
(relative to the magnetic axis) of the propagation direction that differs from 
the tangent to the field line, the condition for waves to propagate into the CFLR is (Appendix): 
\begin{equation}
r_*+s\approx\zeta^2R_{LC}\sin^2(\psi+s/R_{LC}),
\label{eq:condition}
\end{equation}
where $\Psi=\psi+s/R_{LC}$ and $\zeta\leq1$. Note that $\psi$ can be expressed in terms
of $r_*$, $s$ and $\sigma$ using Eq~(\ref{eq:A1}) and (\ref{eq:A6}). This 
condition is shown in Figure~\ref{fig:gc}.  
The condition can easily be satisfied for emission originating
from a large emission radius and a field line close to the last
open field line $\zeta\sim1$.

\subsubsection{Fast pulsars}

The fate of the radio emission that propagates into the CFLR is determined 
by the opacity (\ref{eq:tau3}).  We consider PSR 1937+21 and the Crab pulsar.
For PSR 1937+21, we have the mean flux density $S^{\rm obs}_\nu\approx 240\,{\rm mJy}$ 
at 400 MHz and 16 mJy at 1.4 GHz.  The brightness temperature times
$\Delta\Omega$ at $R_{LC}$ is estimated to be $\Delta\Omega
T_b\sim 10^{26}\,\rm K$ and $5\times10^{23}\, {\rm K}$, respectively.
The maximum depth from the last closed field
lines boundary is $\Delta h_{\rm max}\approx (r_*+s)\Delta\psi=
(r_*+s)^{3/2}/(1-r_*-s)^{1/2}\Delta\zeta\approx
2.5\times10^{-3}R_{\rm LC}$ for $r_*+s=0.5$, $\Delta\zeta=0.01$. For
$P=1.5\, \rm ms$  one has $\Delta h_{\rm max}\approx2\times10^2 \rm m$.
Using (\ref{eq:rcr}) and assuming $B=10^5\,\rm T$ and $\Delta s=0.2R_{LC}$, 
one finds $r_{cr}>R_{LC}$. The opacity at $r\sim R_{LC}$ is $\tau\sim 53|\xi_p|^2\gg 1$ at $\nu=1.4\,
\rm GHz$ and it is even larger at a lower frequency. This implies that
propagation in the CFLR is inhibited. Since both the leading  
and trailing components are visible, the emission must be produced at a radius
where its trailing part does not propagate into the CFLR. From figure~\ref{fig:gc}, 
this radius must be in the range $r_*<0.13R_{LC}$ or $r_*>0.5R_{LC}$. 
Since the LC radius is only $R_{LC}\approx 7.2R_0$ and $r_*>R_0\approx0.14R_{LC}$, 
the only permitted radial range is $r_*>0.5R_{LC}$. Alternatively, one may assume the radio emission
occurs in a region away from the last closed field line surface and in
this case it propagates only in the OFLR. However, one needs to 
assume a two-pole model to explain the double peak feature.

For the Crab pulsar, the observed flux density is $S^{\rm obs}_\nu\approx950\,
\rm mJy$ at 400 MHz and $S^{\rm obs}_\nu\approx14\,{\rm mJy}$ at
1.4 GHz, corresponding a spectral index $\alpha\sim 3.4$. This gives 
$\Delta\Omega T_b\sim 3\times10^{20}\,\rm K$ for $\nu=1.4\,\rm GHz$ at $r=R_{LC}\approx
1.6\times10^6\,\rm m$ and $r_{cr}\sim 0.5R_{LC}|\xi_p|^{4/7}$. Since the propagation in 
the CFLR is nearly parallel, one has back scattering with $|\xi_p|^2\sim {n''}^2\gg 1$.
Since both the leading and trailing pulses are observed, the emission radius must be 
in the range $r_*<0.13R_{LC}$ or $r_*\geq0.5R_{LC}$. 

\subsection{Constraint on backward emission}

Induced three-wave processes impose a severe constraint on any model involving backward 
emission in the pulsar frame. Backward emission may be relevant for  oscillatory gap models such as
proposed by ~\cite{lmjl05}. In this model, pair cascades occur in an 
oscillatory manner sending secondary particles both forward and backward. This would 
lead to two photon beams propagating in opposite directions. Backward emission is also
proposed for interpretation of specific features of the pulse profile (Dyks et al. 2005).
It is suggested here that the backward propagating waves must pass through the CFLR and 
undergo strong induced scattering. As the zeroth order approximation, one assumes
the photon beam is confined to the magnetic meridian plane and ignoring both aberration
and general relativity, one may estimate the radial distance to the 
crossing point $(r_c,\psi_c)$ from 
$r_c/\cos(\sigma+\psi_*)=r_*/\cos(\sigma-\psi_c)$.  
This gives
\begin{equation}
r_c\sim {\psi_*\over{3\psi_*+2\psi_c}}r_*,
\label{eq:rc}
\end{equation}
where $\psi_*\ll1$ is assumed, and $\pi/2-\sigma\sim 3\psi_*/2$.
The polar angles $\psi_c$ and $\psi_*$ can be expressed in terms of the
relevant radii: $\psi_c\sim (r_c/R_{LC})^{1/2}$ and $\psi_*\sim (r_*/R_{LC})^{1/2}$.
On substituting them into (\ref{eq:rc}), one finds the solution $r_c\sim r_*/4$.
When aberration is included, the propagation path of the backward emission
bends toward the star in the corotating frame. A more rigorous treatment is
outlined in Appendix. Since for fast rotating young pulsars 
or fast millisecond pulsars, the critical radius $r_{cr}$
is close to or larger than the LC radius, propagation of the backward emission in
the CFLR is not allowed. For example, as shown in Sec 4.3, the Crab pulsar has 
$r_{cr}>0.5R_{LC}$ for $n''>1$. For the backward emission not to pass through the
opaque region ($r<r_{cr}$), the emission radius must satisfy $r_*>4r_{cr}>2R_{LC}$.
This implies that no backward emission can pass through the CFLR. 

For a typical pulsar, the critical radius can be smaller than the LC radius.
Pulsars satisfy this condition ($r_{cr}<R_{LC}$) must have a period 
\begin{eqnarray}
P&>&(0.1\,{\rm s}) \left({\Delta\Omega T_b\over 3\times10^{22}\,{\rm K}}\right)^{1/5}
\left({\nu\over1\,{\rm GHz}}\right)^{-1/5}\left({B\over
10^8\,{\rm T}}\right)^{3/10}\nonumber\\
&&\times\left({n''\over50}\right)^{2/5}.
\end{eqnarray}
Backward emission can propagate through the region $r_{cr}<r<R_{LC}$, provided that
the emission is produced close to the LC, $4r_{cr}<r_*<R_{LC}$. 
We consider PSR B0950+08 as an example.
It has a period $P=0.253\,\rm s$ and a magnetic field $B\approx 10^8\,{\rm T}$.
We estimate $\Delta \Omega T_b\sim 10^{17}\,{\rm K}$ and
$r_{cr}\sim 0.014R_{LC}{n''}^{4/7}$. Backward emission can propagate
through the CFLR if the emission region is at $r_*>0.056R_{LC}{n''}^{4/7}$.
Note that $n''$ can be much larger than unity. For $n''=10$, one has
$r_*>0.2R_{LC}$. Thus even for a typical pulsar, propagation of 
radio emission in the CFLR is strongly constrained and is allowed
only in the region near the LC. 

\section{Conclusions}

We consider the effect of induced three-wave turbulence on
propagation of the radio emission in a pulsar magnetosphere. 
(As remarked in Sec. 2.5, this process is sometimes referred to, inappropriately
we believe, as `stimulated Raman' or `induced Raman' scattering.)
The low frequency wave is either an electrostatic plasma wave 
propagating along magnetic field lines or an Alfv\'en wave propagating 
nearly along the magnetic field. We discuss both cases of propagation 
in the OFLR and CFLR. In the polar (OFLR)  region, the induced 
processes strongly depend on the bulk velocity of the plasma and on the 
pair multiplcity ($M$).  Since the scattering rate (cf. Eq. \ref{eq:Gamma3}) 
decreases for increasing $\gamma$ and $M$, the induced three-wave scattering
effect on the propagation is suppressed due to the 
relativistic bulk flow ($\gamma\gg1$) and $M\gg1$. 
Assuming a modest bulk Lorentz factor $\gamma\sim10^2$ and a 
multiplicity $M\sim10^2$ for a young pulsar like the Crab pulsar, and
$\gamma\sim 10^5$ and $M\sim 10$ for a fast millisecond pulsar
like PSR 1957+21, we show that propagation of radio emission 
in the OFLR is little affected by the induced three-wave processes. An 
exception is the propagation of giant pulses. If the giant pulses are emitted 
relatively close to the star, they are subject to very strong induced 
processes that prevent them from escaping. The favored region where
induced processes can be effective is the transition region located 
at $r_*/\gamma$ from the source, where the photon beam propagates nearly 
perpendicular to the magnetic field in the rest frame. For giant pulses not 
to be completely destroyed by such processes, their emission region needs to 
be close to the LC. This conclusion is consistent with the conclusion
reached by \cite{rj01} from a different argument, based on the alignment of giant 
pulses with the high energy emission. 
 
Induced three-wave interactions are generally important for propagation 
of radio emission in the CFLR. Because the plasma in this region
is stationary in the corotating frame, with $|\eta|=1$, the processes 
are much more efficient than in the polar region. The induced three-wave processes 
are important throughout the CFLR except for pulsars with a relatively long 
period. We discuss three specific examples where the processes are relevant. 
(1) For the double pulsar J0737-3039B, we find that the processes 
discussed here are not efficient enough to explain the observed eclipse,
due to the very low plasma density of the CFLR of pulsar B which has
a relatively long period. However, the induced three-wave scattering can be effective
if one assumes a density much higher than the GJ density (by a factor of $10^4$).
It is worth noting that a much higher plasma density is  
needed in the synchrotron/cyclotron absorption model as well~\citep{lt05,rg05}. (2) 
Our result imposes a strong constraint on the location 
of the emission region of pulsars with a pulse profile with widely separated leading and
trailing components. There is a particular radial range where the trailing component may be swept into the CLFR
due to aberration. (3) Our result limits the visibility of backward emission.
The backward emission must propagate through the CFLR where
it can be dispersed due to three wave interactions.
For young or fast millisecond pulsars, the radius of the opaque region
(where induced three-wave interactions are important) is comparable with
or larger than the LC radius. In this case, backward emission may not be 
visible. For long period pulsars, the plasma density in the CFLR is 
low and the radius of the opaque region is smaller than the LC.
The backward emission can propagate through the magnetosphere
provided that it is produced sufficiently close to the LC and that
its propagation path is outside the opaque sphere. 

An important approximation made in our discussion is the strong magnetic field limit.  
Although such an approximation excludes the possibility of cyclotron resonance,
our result should be valid for a finite, strong magnetic field. This is because
the relevant frequency of the low frequency waves (Langmuir and Alfv\'en modes) in 
the induced three-wave interactions considered here is much lower than the cyclotron 
frequency throughout the pulsar magnetosphere. 

\appendix

\section{Photon beam in the corotating frame}

In the non-rotating frame, a photon beam is assumed to be emitted
at $(x_*,y_*)$ in the field line direction and to propagate along a 
straight line described by
\begin{equation}
x(s)=x_*\pm s\cos\sigma,\quad\quad\quad y(s)=y_*\pm s\sin\sigma,
\label{eq:A1}
\end{equation}
where $s=ct/R_{\rm LC}=\Omega t$ is the distance  
propagated in time $t$, $\sigma$ is the slope of the photon beam,
and $\pm$ correspond respectively to the forward and backward emission. 
Here, all the relevant radii, distances are in units of $R_{\rm LC}$. 
We consider an orthogonal rotator with a dipole field given by
\begin{equation}
x=\zeta^2\sin^3\psi,\quad\quad\quad y=\zeta^2\cos\psi\sin^2\psi,
\end{equation}
where 
\begin{equation}
\zeta={\sin\psi_d\over\sin\psi_B}
\end{equation}
is a parameter that specifies a particular field line, where $\psi_d$ is
the half-opening angle of the polar cap,  $\psi_B$ is the colatitude
angle. The open field lines correspond to $\psi_B\leq\psi_d$, that is
$\zeta\geq1$. For a nonrotating dipole, one must have
\begin{equation}
\tan\sigma={3\cos^2\psi_*-1\over3\sin\psi_*\cos\psi_*}.
\label{eq:A2}
\end{equation}

To include the rotation, we consider a corotating frame in which the propagation path
is described by
\begin{eqnarray}
X(s)&=&R(s)\sin[\Psi(s)],\nonumber\\
Y(s)&=&R(s)\cos[\Psi(s)],
\label{eq:A3}
\end{eqnarray}
where $R(s)=[X(s)^2+Y(s)^2]^{1/2}=r(s)$, $\Psi(s)=\psi(s)+s$, and
\begin{eqnarray}
r(s)&=&[x(s)^2+y(s)^2]^{1/2},\nonumber\\
\psi(s)&=&\arctan[x(s)/y(s)].
\label{eq:A4}
\end{eqnarray}
The colatitude angle, radial distance of the emission point are
$\Psi(0)\equiv\Psi_*=\psi(0)\equiv\psi_*$, $R(0)=r(0)\equiv r_*$.
The tangent  is
\begin{eqnarray}
{dY\over dX}&=&{Y'(s)\over X'(s)}=
{r'\cos(\psi+s)-r(1+\psi')\sin(\psi+s)\over
r'\sin(\psi+s)+r(1+\psi')\cos(\psi+s)}\nonumber\\
&=&
{\sin(\sigma-s)\mp r\sin(\psi+s)
\over\cos(\sigma-s)\pm r\cos(\psi+s)}.
\label{eq:A5}
\end{eqnarray}
We use
\begin{eqnarray}
r'(s)&=&{x(s)x'(s)+y(s)y'(s)\over r(s)}\nonumber\\
&=&\pm {x(s)\cos\sigma+y(s)\sin\sigma\over
r(s)}=\pm \sin(\sigma+\psi),\\
\psi'(s)&=&{x'(s)y(s)-y'(s)x(s)\over
r(s)^2}\nonumber\\
&=&\pm {y(s)\cos\sigma-x(s)\sin\sigma\over r(s)^2}=
\pm {\cos(\sigma+\psi)\over
r(s)}.
\label{eq:A6}
\end{eqnarray}
 At $s=0$ we have
\begin{equation}
{dY\over dX}\biggr|_{s=0}={\sin\sigma\mp r_*\sin\psi_*
\over
\cos\sigma\pm r_*\cos\psi_*}.
\label{eq:A7}
\end{equation}
Since the derivation of (\ref{eq:A7}) is in the approximation
to the order $\beta=v/c$, it can also be derived by a Galilean transformation. 
Assuming $dY/dX|_{s=0}=\tan\bar{\sigma}$, i.e. the tangent of the beam at
$s=0$ in the corotating frame, we obtain
\begin{equation}
\sin(\sigma-\bar{\sigma})=\pm r_*\sin(\bar{\sigma}+\psi_*).
\label{eq:A8}
\end{equation}
The $+$ sign corresponds to forward emission and $-$ to backward emission.
Due to the aberration effect, the photon beam must make an angle $\sigma\neq\bar{\sigma}$
to the field line in the observer's frame so that it is tangent to the 
field line in the corotating frame. We assume the magnetic field is a
static dipole in the corotating frame:
\begin{equation}
X=\zeta^2\sin^3\Psi,\quad\quad\quad Y=\zeta^2\cos\Psi\sin^2\Psi,
\label{eq:A9}
\end{equation}
where $\zeta=\sin\Psi_d/\sin\Psi_B$.
The tangent to the field line at $s=0$ is
\begin{equation}
\tan\bar{\sigma}={3\cos^2\Psi_*-1\over3\sin\Psi_*\cos\Psi_*}.
\label{eq:A10}
\end{equation}
In the small angle approximation $\psi_*\ll1$, we have $\pi/2-\tilde{\sigma}\approx
3\psi_*/2$ and then (\ref{eq:A8}) reduces to 
\begin{equation}
\sigma-\tilde{\sigma}\approx \pm r_*. 
\end{equation}
The photon beam direction at $r_*$ in the observer's frame advances in the rotation 
direction by $r_*$. 

For the trailing component, the condition for the forward propagating photon 
beam to cross the closed field lines with $\zeta\leq1$ is $R=\zeta^2\sin^2\Psi$, 
where $R(s)=r(s)=[x(s)^2+y(s)^2]^{1/2}$, that is
\begin{equation}
\left[(r^2_*+s^2\pm 2r_*s\sin(\psi_*+\sigma)\right]^{1/2}=\zeta^2\sin^2(\psi+s),
\label{eq:A11}
\end{equation}
where $\tan\psi=(r_*\sin\psi_*+s\cos\sigma)/(r_*\cos\psi_*+s\sin\sigma)$.
Note that the $-$ sign corresponds to the backward emission propagating into
the CLFR on the leading side. The condition is therefore that there exists at least a nontrivial 
solution to Eq. (\ref{eq:A11}) for $s$ and $\zeta\leq1$ within the LC.
In the case of forward emission, on approximating the left-hand side of Eq. (\ref{eq:A11}) by 
$r_*+s$ and using $R(s)=r(s)$, one obtains the approximate form
(\ref{eq:condition}) after reversing them back to their
dimensional form $r_*\to r_*/R_{Lc}$, $s\to s/R_{LC}$.  Similarly, the condition 
(\ref{eq:A11}) for the backward emission (the $-$ sign) can be written approximately as 
$r_*-s\approx (\psi+s)^2$ in the small angle approximation, where $\psi
\approx (r_*-3s\psi_*/2)/(r_*-s)$. One finds that the 
photon beam enters the CFLR on the leading side at a radius $r_c\approx r_*/4$.

\end{document}